\documentclass[aps,prd,preprint,nofootinbib,superscriptaddress,tightenlines]{revtex4}
\usepackage{amsfonts}
\usepackage{mathrsfs}
\usepackage{graphicx}
\usepackage{amsmath}
\usepackage{amssymb}
\usepackage{multirow}
\usepackage{subfigure}
\usepackage{epsfig}
\usepackage{graphicx}
\usepackage{booktabs}
\usepackage{array}
\usepackage{tabularx}
\usepackage{slashed}
\usepackage{ulem}
\usepackage{bm}

\newcommand{\bqa}{\begin{eqnarray}}
\newcommand{\eqa}{\end{eqnarray}}
\newcommand{\beq}{\begin{equation}}
\newcommand{\eeq}{\end{equation}}
\graphicspath{{fig/}{dia/}} \DeclareGraphicsExtensions{.eps}
\begin{document}
\baselineskip 20pt
\title{The exclusive production of a fully heavy tetraquark and a photon in electron-positron collision}

\author{Xiao Liang}
\email{202120290@mail.sdu.edu.cn}
\affiliation{School of Physics and Optoelectronic Engineering, Shandong University of Technology, Zibo, Shandong 255000, China
}

\author{Jun Jiang}
\email{jiangjun87@sdu.edu.cn}
\affiliation{School of Physics, Shandong University, Jinan, Shandong 250100, China}

\author{Shi-Yuan Li}
\email{lishy@sdu.edu.cn}
\affiliation{School of Physics, Shandong University, Jinan, Shandong 250100, China}

\author{Yan-Rui Liu}
\email{yrliu@sdu.edu.cn}
\affiliation{School of Physics, Shandong University, Jinan, Shandong 250100, China}

\author{Zong-Guo Si}
\email{zgsi@sdu.edu.cn}
\affiliation{School of Physics, Shandong University, Jinan, Shandong 250100, China}

\begin{abstract}

The exclusive production of fully heavy tetraquark ($T(bb\bar{b}\bar{b})$, $T(cc\bar{c}\bar{c})$ and $T(bc\bar{b}\bar{c})$) in association with a hard photon in electron-positron collisions is calculated within the framework of non-relativistic QCD. 
Both inner structures of molecule-like state and compact state with $J=0,1,2$ for the fully heavy tetraquark are discussed. 
We find that it is dismal to observe any fully heavy tetraquarks in either the compact configuration or the molecule-like configuration through such exclusive processes at either Belle II or future Z factories like CEPC and FCC-ee.

\vspace {2mm} 
\noindent {Keywords: Fully heavy tetraquark, NRQCD, exclusive production }
\end{abstract}

\maketitle

\section{INTRODUCTION}
\label{sec:introduction}

Exotic hadrons, distinct from the conventional quark-antiquark or three-quark states of conventional hadrons, have captured extensive attention. Comprehensive investigations into their characteristics will provide an ultimate assessment of the fundamental theory and enhance our understanding of non-perturbative QCD. Since the first tetraquark candidate $\chi_{c1}(3872)$ was discovered by the Belle collaboration in 2003 \cite{Belle:2003nnu}, approximately 50 exotic hadron candidates have been identified in experiments worldwide to date. This has marked the dawn of a new era in the study of heavy exotic hadrons. In particular, the LHCb collaboration reported evidence of a narrow structure near $6.9\ {\rm GeV}$ in the di-$J/\psi$ channel in 2020 \cite{LHCb:2020bwg}, and this was subsequently confirmed by both the ATLAS \cite{ATLAS:2023bft} and CMS collaborations \cite{CMS:2023owd}. This structure, known as $X(6900)$, is widely regarded as a highly probable candidate for a fully charmed tetraquark $T_{4c}$. This discovery has subsequently spurred in-depth theoretical research into the characteristics of the relatives of fully heavy tetraquarks.

Given the non-relativistic nature of the interior of fully heavy tetraquarks, the non-relativistic QCD (NRQCD) factorization framework \cite{Bodwin:1994jh,Petrelli:1997ge} can be employed to describe the production process. This framework, which factorizes perturbative and nonperturbative effects and has been widely used to characterize various quarkonium and diquark production and decay processes, is well-suited for this task. Under the NRQCD factorization formalism, the production rate is decomposed into short-distance coefficients and long-distance matrix elements. The short-distance coefficients correspond to the hard creation of heavy quark pairs and can be calculated perturbatively using Feynman diagram methods. In contrast, the long-distance matrix elements represent the nonperturbative hadronization of diquarks into fully heavy tetraquarks. Two dominant configurations have been proposed to describe fully heavy tetraquarks: the so-called compact multiquark states and molecular states. In this work, we apply the NRQCD framework to calculate the production of fully heavy tetraquarks in both configurations.

Up to now, the mass spectra and the production and decay of fully heavy tetraquarks have been comprehensively investigated by employing various models and techniques in the existing literature. For instance, the mass spectra and decay modes of fully heavy tetraquark have been explored within the quark potential model \cite{Bai:2016int,Berezhnoy:2011xn,Becchi:2020uvq,Debastiani:2017msn,Lu:2020cns,Mutuk:2021hmi,Mutuk:2022nkw}, and in the context of QCD sum rules \cite{Wang:2018poa,Chen:2020xwe,Wang:2020ols,Agaev:2024qbh,Agaev:2024wvp,Chen:2024bpz}. 
Recently, many groups have explored the properties of the production and decay processes of fully heavy tetraquarks within the NRQCD factorization approach \cite{Feng:2020riv,Zhang:2023ffe,Feng:2020qee,Huang:2021vtb,Zhu:2020xni,Feng:2023agq,Zhang:2020hoh}. 
Among them, Zhang et al. discussed the decay of fully charm tetraquark $T_{4c}$ into both light hadrons and open charms within both compact state and molecular state \cite{Zhang:2023ffe},
Zhang and Ma studied the inclusive production of di-$J/\Psi$ resonances at LHC \cite{Zhang:2020hoh},
and Zhu studied the inclusive production of fully heavy tetraquarks in the $J/\Psi$-pair mass spectrum
using proton-proton collision data by the LHCb collaboration \cite{Zhu:2020xni}. 
Especially, Zhang and Ma extended the NRQCD framework to explore the fully heavy tetraquak states \cite{Zhang:2023ffe,Zhang:2020hoh}.
Feng et al. also proposed a novel NRQCD factorization theorem for the production of the fully heavy tetraquarks, and conducted a series of studies on fully charm tetraquark $T_{4c}$ in the compact state \cite{Feng:2020riv,Feng:2020qee,Feng:2023agq,Huang:2021vtb}, including in $pp$ collision \cite{Feng:2020riv,Feng:2023agq} and at an electron-positron collider for $1^{+-}$ states \cite{Huang:2021vtb}, $0^{++}$ and $2^{++}$ states \cite{Feng:2020qee}.
Especially, in Ref. \cite{Feng:2020qee}, they studied the compact fully charm tetraquark at the B factory, and we compare parts of our results with theirs in this paper and obtained the same results. 
In addition, the production of $T_{4c}$ from light quark fragmentation \cite{Bai:2024flh}, charm quark fragmentation \cite{Bai:2024ezn}, the photoproduction of $T_{4c}$ at electron-ion colliders \cite{Feng:2023ghc}, the electromagnetic and hadronic decay of fully heavy tetraquarks \cite{Sang:2023ncm}, and $T_{4c}$ production at LHC based on a factorization scheme which employs the tetraquark four-body wave function \cite{Belov:2024qyi} are explored.

In this manuscript, within the NRQCD factorization framework, we calculate the exclusive production of $S$-wave fully heavy tetraquarks—$T(bb\bar{b}\bar{b})$, $T(cc\bar{c}\bar{c})$, and $T(bc\bar{b}\bar{c})$—with total spin $J=0,1,2$ in association with a photon in electron-positron collisions, and both molecular-like and compact configurations of these tetraquarks are discussed. To enhance the utility of our work for future experimental studies, we present numerical results for the production cross sections and event counts at both ultra-high-luminosity B factories (e.g., the recently reactivated Belle II experiment) and $Z$ factories (e.g., the Circular Electron Positron Collider (CEPC) \cite{CEPCStudyGroup:2018ghi} and the first stage of the Future Circular Collider (FCC-ee) \cite{Agapov:2022bhm}). Due to the constraint imposed by the center-of-mass energy $\sqrt{s}$ of the electron-positron system, only $T(cc\bar{c}\bar{c})$ can be produced at B factories.

The rest of the paper is organized as follows. 
In section \ref{sec:formulation},  
the main theoretical framework of fully heavy tetraquarks of  molecule-like state and compact state is presented, respectively.
Meanwhile, some technical particulars utilized in the calculation are introduced.
In section \ref{sec:data}, the numerical results at both B factory and Z factory will be presented.
The section \ref{sec:summary} is reserved for a summary.

\section{FORMULATION}
\label{sec:formulation}
\subsection{NRQCD factorization for fully heavy tetraquark}

\begin{figure}[!thbp]
    \centering
    \includegraphics[width=0.8\textwidth]{{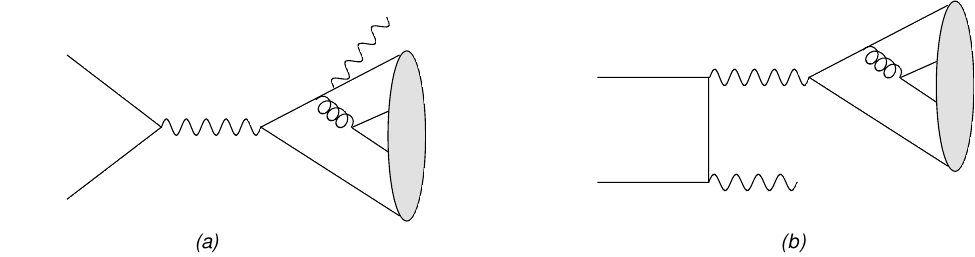}}
    \caption{Typical Feynman diagrams for $e^{+}+e^{-}\to \gamma^{*}/Z^{0}\to T+\gamma$. Diagram (a) represents the $s$-channel diagrams and (b) designates the $t$-channel diagrams.}
    \label{diagram}
\end{figure}

The typical Feynman diagrams for the process $e^{+}+e^{-}\to \gamma^{*}/Z^{0}\to T+\gamma$ are presented in Fig. \ref{diagram}, where $T$ denotes fully heavy tetraquarks with constituent quarks $bb\bar{b}\bar{b}$, $cc\bar{c}\bar{c}$, and $bc\bar{b}\bar{c}$.
Since we only consider $S$-wave tetraquarks, the $P$-parity of $T$ is always positive. Consequently, Fig. \ref{diagram} (b) is excluded by $P$-parity conservation, as the photon propagator has negative $P$-parity. For processes involving the $Z^0$ propagator, we omit Fig. \ref{diagram} (b) at $Z$ factories: the $Z^0$ decay primarily produces $T$ via the topology shown in Fig. \ref{diagram} (a), which yields a much larger production rate.
Fig. \ref{diagram} (a) comprises 40 Feynman diagrams for $T(bb\bar{b}\bar{b})$ or $T(cc\bar{c}\bar{c})$, and 20 diagrams for $T(bc\bar{b}\bar{c})$.

Regarding the internal structures of fully heavy tetraquarks $T(bb\bar{b}\bar{b})$, $T(cc\bar{c}\bar{c})$ and $T(bc\bar{b}\bar{c})$, we will discuss two kinds of situations, molecule-like tetraquarks which are 
composed of double quarkonia, and compact tetraquarks which are composed of a diquark and an antidiquark. Under the NRQCD factorization framework \cite{Bodwin:1994jh,Petrelli:1997ge}, the cross sections for the production of fully heavy tetraquarks in electron-positron collision can be factored into the short-distance coefficients and the long-distance matrix elements. Therefore, for the molecule-like state and the compact tetraquark state, the cross section can be written as \cite{Bodwin:1994jh}
\begin{equation}
     \sigma(e^{+}+e^{-}\to T_{M/C}+\gamma)=  \hat\sigma(e^{+}+e^{-} \to Q_1 Q_2 \bar{Q}_1\bar{Q}_2 +\gamma) {\langle O_{\overline{{\mathbf{c}}} , \mathbf{c}}^J (T_{M/C}) \rangle}\ .     
    \label{cross section}
\end{equation}
Here, the subscripts $M$ and $C$ stand for the molecule-like tetraquark and the compact tetraquark, respectively.
$\hat\sigma$ describes the short-distance coefficients, and the long-distance matrix elements ${\langle O_{\overline{{\mathbf{c}}} , \mathbf{c}}^J (T_{M/C}) \rangle}$ stand for the transition probability of four heavy quarks to $T_{M/C}$ accordingly, where $\mathbf{c},\,\mathbf{\bar{c}}$ are for the colors of constituent quarkonia or (anti-)diquarks, and $J$ for the total spin of tetraquark. 
Within the NRQCD framework, the production operators ${ O_{\overline{{\mathbf{c}}} , \mathbf{c}}^J (T_{M/C}) }$ for the molecule-like state and the compact tetraquark state are defined via \cite{Bodwin:1994jh,Bai:2024ezn}
\begin{equation}
    O_{\overline{\mathbf{c}} , \mathbf{c}}^{J}(T_{M/C}) =\mathcal{O}_{\overline{\mathbf{c}} \otimes \mathbf{c}}^{J} \sum_X\left|T_{M/C}^J+X\right\rangle\left\langle T_{M/C}^J+X\right| \mathcal{O}_{\overline{\mathbf{c}} \otimes \mathbf{c}}^{J \dagger},
\end{equation}
where $X$ stands for all other final-state particles, and the colorless four-quark NRQCD operators $\mathcal{O}_{\overline{\mathbf{c}} \otimes \mathbf{c}}^{J}$ have the following explicit definitions \cite{Bodwin:1994jh,Feng:2020riv,Feng:2020qee,Bai:2024ezn},
\begin{equation}
\begin{aligned}
& {\cal O}_{{\mathbf{1}} \otimes \mathbf{1}}^{0}(T_{M}) = \left[\psi_a^{T} \left(i \sigma^2\right)\sigma^i \psi_{\bar{a}}\right] \left[\chi_b^\dagger \sigma^i \left(i \sigma^2\right)  \chi_{\bar{b}}^*\right], \\
& O_{{\mathbf{1}} \otimes \mathbf{1}}^{2, ij} (T_{M}) = \left[\psi_a^{T}  \left(i \sigma^2\right) \sigma^m \psi_{\bar{a}} \right]\left[\chi_b^\dagger \sigma^n \left(i \sigma^2\right)  \chi_{\bar{b}}^*\right] \Gamma^{ij ; m n}; \\
& {\cal O}_{\overline{\mathbf{3}} \otimes \mathbf{3}}^{0}(T_{C}) = \left[\psi_a^{T} \left(i \sigma^2\right)\sigma^i \psi_b\right] \left[\chi_c^\dagger \sigma^i \left(i \sigma^2\right)  \chi_d^*\right] C_{\overline{\mathbf{3}} \otimes \mathbf{3}}^{a b ; c d}, \\
& O_{\overline{\mathbf{3}} \otimes \mathbf{3}}^{1, i }(T_{C}) = \left[\psi_a^{T}  \left(i \sigma^2\right) \sigma^j \psi_b \right]\left[\chi_c^\dagger \sigma^k \left(i \sigma^2\right)  \chi_d^*\right] \epsilon^{ijk} C_{\overline{\mathbf{3}} \otimes \mathbf{3}}^{a b ; c d}, \\
& O_{\overline{\mathbf{3}} \otimes \mathbf{3}}^{2, ij}(T_{C}) = \left[\psi_a^{T}  \left(i \sigma^2\right) \sigma^m \psi_b \right]\left[\chi_c^\dagger \sigma^n \left(i \sigma^2\right)  \chi_d^*\right] \Gamma^{ij ; m n} C_{\overline{\mathbf{3}} \otimes \mathbf{3}}^{a b ; c d}, \\
& O_{\mathbf{6} \otimes \overline{\mathbf{6}}}^{ 0} (T_{C}) = \left[\psi_a^{T}\left(i \sigma^2\right) \psi_b\right]\left[\chi_c^\dagger \left(i \sigma^2\right) \chi_d^* \right] C_{\mathbf{6} \otimes \overline{\mathbf{6}}}^{a b ; c d},
\end{aligned}
\label{eq:operators}
\end{equation}
where $\psi^{\dagger}$ and $\chi$ are the Dirac spinor fields to create heavy quark $Q$ and antiquark $\bar{Q}$, $\sigma^i$ ($i=1,2,3$) are Pauli matrices, $\epsilon^{ijk}$ is the antisymmetric tensor, the tensor $\Gamma^{ij ; m n} \equiv g^{im}g^{jn} + g^{in}g^{jm}-\frac{2}{3}g^{ij}g^{mn}$, and $C_{\mathbf{c} \otimes \overline{\mathbf{c}}}^{a b ; c d}$ are the color factors of tetraquark with $a,b,c,d$ being color indices of quarks.
Here in this work, we omit the contribution of $\mathbf{8} \otimes \overline{\mathbf{8}}$ color configuration for the molecular-like state because the color-octet $(Q\bar{Q})^{\mathbf{8}}$ pair is suppressed by $v^2$ in comparison with the color-singlet $(Q\bar{Q})^{\mathbf{1}}$ pair in the NRQCD velocity expansion, so the $\mathbf{8} \otimes \overline{\mathbf{8}}$ color configuration is suppressed by $v^4$. And usually the contribution of color-octet $(Q\bar{Q})^{\mathbf{8}}$ pair is negligible at small transverse momentum regions (like the B factory in this work) because the fragmentation of $g \to (Q\bar{Q})^{\mathbf{8}}$ would become important only at large transverse momentum regions which can be reached at LHC.

It should be noted that we adopt a two-step transition scheme for the formation of fully heavy tetraquarks from four heavy quarks. First, a $Q\bar{Q}$ pair combines into a colorless quarkonium, or two (anti-)quarks ($QQ$ or $\bar{Q}\bar{Q}$) combine into a colored (anti-)diquark. Second, two quarkonia are bound into a molecular-like state, or a diquark-antidiquark pair is bound into a compact state.
However, Ref. \cite{Bai:2016int} points out that once the distance between the diquark and antidiquark becomes comparable to the separation of their constituent quarks, the diquark approximation breaks down, and a full four-body treatment becomes necessary.
Four-body Schr$\ddot{\mathrm{o}}$dinger wave functions for fully charmed tetraquarks have been extensively studied using various potential models \cite{Lu:2020cns,Zhao:2020nwy,liu:2020eha,Yu:2022lak,Wang:2019rdo}. For instance, Ref. \cite{Bai:2024ezn} discusses the values of long-distance matrix elements under five distinct potential models. 

In the rest of this section, we will present the explicit framework for the calculation of both molecule-like tetraquark and compact tetraquark in the two-step transition scheme.
We follow the extended NRQCD framework in Refs. \cite{Zhang:2023ffe,Zhang:2020hoh} by Zhang and Ma to explore the production of fully heavy tetraquak states.
We use ``meson'' to indicate either the quarkonium or the (anti-)diquark for convenience.
The creation operator of a tetraquark can be written as
\begin{equation}
    \begin{split}
a_{T_{M/C}}^\dagger&(P,J,J_z)=\sum_{\lambda_1\lambda_2} C_{J J_z}^{\lambda_1\lambda_2} \int\frac{\mathrm{d}^3p_1}{(2\pi)^32p_1^0}\frac{\mathrm{d}^3p_2}{(2\pi)^32p_2^0}  \\
&\times(2\pi)^32P^0\delta^3(P-p_1-p_2)\tilde{\psi}_{M/C}(P,p_1,p_2)a_{meson1}^\dagger(p_1,\lambda_1)a_{meson2}^\dagger(p_2,\lambda_2).  
\end{split} \label{a+T}
\end{equation}
Here, $C_{J J_z}^{\lambda_1\lambda_2}$ is the Clebsch-Gordan coefficient. 
$\tilde{\psi}_{M/C}(P,p_1,p_2)$ is the spatial wave function of tetraquark $T$, where $P$, $p_1$ and $p_2$ are the momenta of tetraquark and constituent mesons, respectively. $a_{meson1}^\dagger(p_1,\lambda_1)$ and $a_{meson2}^\dagger(p_2,\lambda_2)$ are the creation operators of constituent mesons with momenta $p_1$, $p_2$ and polarization $\lambda_1$, $\lambda_2$. 
We define $P=p_1+p_2$ and $q=(p_2-p_1)/2$, and the phase space integration reduces to
\begin{equation}
    \begin{split}
\frac{\mathrm{d}^3p_1}{(2\pi)^32p_1^0}\frac{\mathrm{d}^3p_2}{(2\pi)^32p_2^0}&=\frac{\mathrm{d}^4p_1}{(2\pi)^3}\frac{\mathrm{d}^4p_2}{(2\pi)^3}\delta(p_1^2-m^2)\delta(p_2^2-m^2)\theta(p_1^0)\theta(p_2^0)  \\
&=\frac{\mathrm{d}^4P}{(2\pi)^3}\frac{\mathrm{d}^4q}{(2\pi)^3}\delta(P^2-4(m^2-q^2))\delta(\frac{P}{2}\cdot q)\theta(\frac{P^0}{2}+q^0)\theta(\frac{P^0}{2}-q^0)  \\
&=\frac{\mathrm{d}^3P}{(2\pi)^32P^0}\frac{\mathrm{d}^3q}{(2\pi)^3}\frac{2P^0}{(P^0)^2- 4(q^0)^2},
\end{split} \label{phase space}
\end{equation}
with $m$ being the masses of the constituent mesons. 
So that Eq. (\ref{a+T}) becomes
\begin{equation}
    \begin{split}
a_{T_{M/C}}^\dagger(P,J,J_z)&=\sum_{\lambda_1\lambda_2}C_{JJ_z}^{\lambda_1\lambda_2}\int\frac{\mathrm{d}^3q}{(2\pi)^3}\frac{2P^0}{(P^0)^2-4(q^0)^2}  \\
&\times\tilde{\psi}_{M/C}(q)a_{meson1}^\dagger(\frac{P}{2}+q,\lambda_1)a_{meson2}^\dagger(\frac{P}{2}-q,\lambda_2),
\end{split} \label{a+T2}
\end{equation}
where $\tilde{\psi}_{M/C}(P,p_1,p_2)$ becomes $\tilde{\psi}_{M/C}(q)$ which depends on $q$ only.
Here, the creation operators of constituent mesons in tetraquark $T$ are normalized
\begin{equation}
    \begin{split}
\Big[a_{meson}(p',\lambda'),a_{meson}^\dagger(p,\lambda)\Big]=(2\pi)^32p^0\delta^3(\bm{p}-\bm{p'})\delta_{\lambda\lambda'}.
\end{split} \label{a+mam}
\end{equation}
Meanwhile, we need the similar normalization for the tetraquark's creation operator 
\begin{equation}
    \begin{split}
\Big[a_{T_{M/C}}(P',J',J_z'),a_{T_{M/C}}^\dagger(P,J,J_z)\Big]=(2\pi)^32P^0\delta^3(\bm{P}-\bm{P'})\delta_{JJ'}\delta_{J_zJ_z'}.
\end{split} \label{a+TaT}
\end{equation}
Combining Eqs. (\ref{a+T2}), (\ref{a+mam}) and (\ref{a+TaT}), we have
\begin{equation}
    \begin{split}
\int\frac{\mathrm{d}^3q}{(2\pi)^3}\frac{2P^0}{(P^0)^2-4(q^0)^2}\big|\tilde{\psi}_{M/C}(q)\big|^2=1. 
\end{split} \label{phase space2}
\end{equation}
Specifically, in the tetraquark rest frame, $P^0=M$, $|\Vec{P}|=0$, $q^0=0$ and $|\Vec{q}|=\sqrt{M^2/4-m^2}$ with $M$ being the mass of the tetraquark, employing which, Eq. (\ref{phase space2}) reduces to
\begin{equation}
    \begin{split}
\int\frac{\mathrm{d}^3q}{(2\pi)^3}\frac{2}{M}\big|\tilde{\psi}_{M/C}(q)\big|^2=1. 
\end{split} \label{rest frame}
\end{equation}
Noting that if the constituent mesons in tetraquark are identical, an additional factor 2 is needed due to the symmetry of interchanging the two mesons.
Then we define the normalized wave functions ${\psi}_{M/C}(q)$ and ${\Psi}_{M/C}(x)$, 
\begin{equation}
    \begin{split}
\int \frac{\mathrm{d}^3q}{(2\pi)^3} \big|{\psi}_{M/C}(q)\big|^2 = 1 =\int \frac{\mathrm{d}^3x}{(2\pi)^3} \big|{\Psi}_{M/C}(x)\big|^2, 
\end{split} \label{normalized wave function}
\end{equation}
where ${\psi}_{M/C}(q)$ and ${\Psi}_{M/C}(x)$ are the wave functions of tetraquark in momentum representation and in coordinate representation, respectively. 
In comparison with Eq. (\ref{rest frame}), we can obtain
\begin{equation}
    \begin{split}
\tilde{\psi}_{M/C}(q)= \frac{\sqrt{M}}{\sqrt{2}}{\psi}_{M/C}(q).
\end{split} \label{normalized wave function2}
\end{equation}

Hence, the S-wave tetraquark consisting of two constituent mesons can be expressed with the creation operator as
\begin{equation}
    \begin{split}
|T_{M/C}(P,J,J_z)\rangle&=a_{T_{M/C}}^\dagger(P,J,J_z)|0\rangle  \\
&=\sum_{\lambda_1\lambda_2}C_{JJ_z}^{\lambda_1\lambda_2}\int\frac{\mathrm{d}^3q}{(2\pi)^3}\frac{2P_0}{(P^0)^2-4(q^0)^2}\tilde{\psi}_{M/C}(q)  \\&
\times a_{meson1}^\dagger(\frac{P}{2}+q,\lambda_1)a_{meson2}^\dagger(\frac{P}{2}-q,\lambda_2)|0\rangle,
\end{split}
    \label{amplitude}
\end{equation}
which, in the tetraquark rest frame, can be simplified as
\begin{equation}
    \begin{split}
&|T_{M/C}(P,J,J_z)\rangle=a_{T_{M/C}}^\dagger(P,J,J_z)|0\rangle  \\
&=\sum_{\lambda_1\lambda_2}C_{JJ_z}^{\lambda_1\lambda_2}\int\frac{\mathrm{d}^3q}{(2\pi)^3}\frac{2}{M}\tilde{\psi}_{M/C}(q)  
a_{meson1}^\dagger(\frac{P}{2}+q,\lambda_1)a_{meson2}^\dagger(\frac{P}{2}-q,\lambda_2)|0\rangle \\
&=\sum_{\lambda_1\lambda_2}C_{JJ_z}^{\lambda_1\lambda_2}\int\frac{\mathrm{d}^3q}{(2\pi)^3}\frac{2}{M}\frac{\sqrt{M}}{\sqrt{2}}{\psi}_{M/C}(q)  
a_{meson1}^\dagger(\frac{P}{2}+q,\lambda_1)a_{meson2}^\dagger(\frac{P}{2}-q,\lambda_2)|0\rangle\\
&=\sum_{\lambda_1\lambda_2}C_{JJ_z}^{\lambda_1\lambda_2}\int\frac{\mathrm{d}^3x}{(2\pi)^3}\frac{\sqrt{2}}{\sqrt{M}}{\Psi}_{M/C}(x)  
a_{meson1}^\dagger(\frac{P}{2}+q,\lambda_1)a_{meson2}^\dagger(\frac{P}{2}-q,\lambda_2)|0\rangle\\
&\approx \sum_{\lambda_1\lambda_2}C_{JJ_z}^{\lambda_1\lambda_2}\frac{\sqrt{2}}{\sqrt{M}}{\Psi}_{M/C}(0)  
a_{meson1}^\dagger(\frac{P}{2}+q,\lambda_1)a_{meson2}^\dagger(\frac{P}{2}-q,\lambda_2)|0\rangle
.
\end{split}
    \label{tetraquark amplitude}
\end{equation}
Note that there is an additional factor $\frac{1}{\sqrt{2}}$ in front of the wave function ${\Psi}_{M/C}(0)$ for the identical constituent mesons in the tetraquark. 
We use the ``$\approx$'' in the final step above because the integral of the wave function is expanded in the entire space at the origin and only the leading term is adopted. 
We can clearly see that Eq. (\ref{tetraquark amplitude}) is Lorentz invariant, which means that it is applicable in arbitrary frames of reference when the value of ${\Psi}_{M/C}(0)$ is evaluated. 

In this work, we consider the vector and scalar constituent mesons in the tetraquark. For the vector situation, which means spin triple S-wave mesons here, the summation over the polarizations of constituent mesons is 
\begin{equation}
    \begin{split}
\epsilon_{JJ_z}^{\mu\nu}=\sum_{\lambda_1\lambda_2}C_{JJ_z}^{\lambda_1\lambda_2}\epsilon_{\lambda_1}^\mu(\frac{P}{2})\epsilon_{\lambda_2}^\nu(\frac{P}{2}),
    \end{split}
    \label{summation of polariazations}
\end{equation}
where $\epsilon_{\lambda_1}^\mu(\frac{P}{2})$ and $\epsilon_{\lambda_2}^\nu(\frac{P}{2})$ are the polarizations of two constituent mesons with the relative momentum $q=0$ for the $S$-wave states.
For $J=0,1,2$, we have
\begin{equation}
    \begin{split}
&\epsilon_{00}^{\mu\nu}(\epsilon_{00}^{\mu'\nu'})^*=\frac{1}{3}\Pi^{\mu\nu}\Pi^{\mu'\nu'},  \\ \sum_{J_z}&\epsilon_{1J_z}^{\mu\nu}(\epsilon_{1J_z}^{\mu'\nu'})^*=\frac{1}{2}(\Pi^{\mu\mu'}\Pi^{\nu\nu'}-\Pi^{\mu\nu'}\Pi^{\nu\mu'}),  \\
\sum_{J_z}&\epsilon_{2J_z}^{\mu\nu}(\epsilon_{2J_z}^{\mu'\nu'})^*=\frac{1}{2}(\Pi^{\mu\mu'}\Pi^{\nu\nu'}+\Pi^{\mu\nu'}\Pi^{\nu\mu'})-\frac{1}{3}\Pi^{\mu\nu}\Pi^{\mu'\nu'},
    \end{split}
    \label{summation of polariazations 0,1,2}
\end{equation}
with
\begin{equation}
    \begin{split}
\Pi^{\mu\nu}=-g^{\mu\nu}+\frac{P^\mu P^\nu}{M^2}.
    \end{split}
\end{equation}

\subsection{Production of molecule-like tetraquarks}
 
In the molecular picture, tetraquark states are described as systems composed of two color-neutral objects, specifically quarkonium-quarkonium interactions, and these interactions are mediated by QCD analogue of the van der Waals force, an idea in the light of the deuteron as a bound state of a proton and a neutron \cite{Husken:2024rdk}. 
In this work, inspired by the discovery of di-$J/\psi$ resonance, we choose to consider the $S$-wave color-singlet vector quarkonium states only in the tetraquark, i.e. double $(c\bar{c})[^3S_1]$ states for $T_{M}(cc\bar{c}\bar{c})$, double $(b\bar{b})[^3S_1]$ for $T_M(bb\bar{b}\bar{b})$, and for $T_M(bc\bar{b}\bar{c})$ we also only consider the mass symmetric state $(c\bar{b})[^3S_1]$-$(b\bar{c})[^3S_1]$ for simplification.

The projector for a heavy quark pair $Q\bar{Q}^\prime$ into a color-singlet and spin-triplet quarkonium has the following form \cite{Petrelli:1997ge}
\begin{equation}
    \begin{split}
\Pi_{q_k} = \epsilon_\alpha(q_k) \frac{-\sqrt{M_{k}}}{4 m_{Q_k}m_{Q^\prime_k}}(\slashed{q}_{k2}- m_{Q^\prime_k}) \gamma^\alpha (\slashed{q}_{k1} + m_{Q_k})\otimes\frac{\delta_{ij}} {\sqrt{N_c}}.
\end{split}
\end{equation}
Here, $q_k$, $q_{k1}$ and $q_{k2}$ are the momenta of quarkonium and its constituent heavy quarks, respectively.  Meanwhile, $M_{k}$, $m_{Q_k}$ and $m_{Q^\prime_k}$ are the corresponding masses of them. $\epsilon_\alpha(q_k)$ is the polarization vector of the quarkonium.

In the two-step transition scheme for the four heavy quarks into the molecule-like tetraquark, the long-distance matrix element ${\langle O_{\overline{\mathbf{c}} , \mathbf{c}}^J (T_{M}) \rangle}$ in Eq. \eqref{cross section} can be replaced by two long-distance matrix elements $\langle O(Q\bar{Q}^\prime)\rangle$ for two $S$-wave constituent quarkonia and one long-distance matrix element $\langle O_{T_M}\rangle$ for two quarkonia being bound into the molecule state,
\begin{equation}
\begin{split}
{\langle O_{{\mathbf{1}} , \mathbf{1}}^J (T_{M}) \rangle} &\longrightarrow \langle O(Q\bar{Q}^\prime)\rangle \langle O(Q\bar{Q}^\prime)\rangle \times \langle O_{T_M}\rangle \\
&=|\Psi_{m_1}(0)|^2 |\Psi_{m_2}(0)|^2 \times \frac{2}{M}|\Psi_{M}(0)|^2 \Gamma^{J} \\
&= \frac{1}{(4\pi)^2} \big|R_{m_1}(0)\big|^2 \big|R_{m_2}(0)\big|^2 \times \frac{2}{4\pi M} \big|R_{M}(0)\big|^2 \Gamma^{J},
\end{split}
\label{eq:transition LDME}
\end{equation}
where the subscripts $m_{1,2}$ and $M$ indicate the two constituent quarkonia and molecule-like tetraquark respectively, and $\Gamma^J$ stands for the polarization summation in Eq. \eqref{summation of polariazations 0,1,2}.
In the second line, the long-distance matrix elements are related to the Schr{$\ddot{\mathrm{o}}$}dinger wave function $\Psi(0)$ at origin in potential models, which can be further related to the radial wave function $R(0)$ evaluated at the origin for $S$-wave states in the third line. 
For clarity, the production long-distance matrix elements for color-singlet spin-triplet quarkonium have the explicit definition as \cite{Bodwin:1994jh}
\begin{equation}
    \langle O(Q\bar{Q}^\prime)[^3S_1]^{\mathbf{1}} \rangle \equiv \chi^\dagger \sigma^i \psi \left( \sum_X \left| (Q
\bar{Q}^\prime) + X \rangle \langle (Q\bar{Q}^\prime)+X \right| \right) \psi^\dagger \sigma^i \chi,
\end{equation}

With the equations given above, the differential cross section for the production of molecule-like tetraquark can be simply written as
\begin{equation}
    \begin{split}
        &\mathrm{d}\sigma(e^{+}+e^{-}\to T_M(J)+\gamma) =  \\
        &\mathrm{d}\hat\sigma(e^{+}+e^{-} \to (Q_1\bar{Q}_2)[^3S_1]+(Q_2\bar{Q}_1)[^3S_1]+\gamma) \times \frac{2\big|R_{m_1}(0)\big|^2 \big|R_{m_2}(0)\big|^2 \big|R_M(0)\big|^2}{(4\pi)^3 M}\ , 
    \end{split}
\end{equation} 
where the polarization summation $\Gamma^J$ has been absorbed into the short-distance cross section $\mathrm{d}\hat\sigma$, which has the definition of
\begin{equation}
\begin{split}
    & \mathrm{d}\hat\sigma(e^{+}+e^{-} \to (Q_1\bar{Q}_2)[^3S_1]+(Q_2\bar{Q}_1) = \\ &\frac{1}{4\sqrt{(p_{e^+} \cdot p_{e^-})^2 - m_e^4}} \overline{\sum} |\mathcal{M}((Q_1\bar{Q}_2)[^3S_1]+(Q_2\bar{Q}_1)[^3S_1]+\gamma)|^2 d\Phi_2.
\end{split} 
\end{equation}
where the two-body phase space $d\Phi_2$ is dimensionless, the squared amplitude $|\mathcal{M}|^2$ has the dimension of $[M]^{-8}$.
So $\mathrm{d} \hat\sigma$ has the dimension of $[M]^{-10}$, which is not a true cross section but stands for all short-distance parts.
In addition, in the cases of $T_M(c\bar{c}-c\bar{c})$ and $T_M(b\bar{b}-b\bar{b})$, an additional factor $1/2!$ should be multiplied due to the symmetry of the two identical constituent mesons.

\subsection{Production of compact tetraquarks}

In the compact tetraquark picture, quarks are strongly bound together via direct strong interactions. A simple model is the diquark antidiquark configuration.
A pair of quarks forms a diquark, and then interacts with the antidiquark formed by the antiquark pair. 
The strong interactions between diquark and antidiquark effectively are the same as those between antiquark and quark in quarkonia \cite{Husken:2024rdk}. 
In this work, we consider both scalar and vector diquark, i.e. $(cc)[n]$-$(\bar{c}\bar{c})[n]$, $(bb)[n]$-$(\bar{b}\bar{b})[n]$ and $(bc)[n]$-$(\bar{b}\bar{c})[n]$ where $[n]$ represents $[^1S_0]$ or $[^3S_1]$ states. 

The amplitudes for the production of diquarks can be related to those for the production of quarkonia.
First, to produce a diquark $( QQ' )$, one of the fermion lines of $Q\bar{Q}$ (or $Q'\bar{Q'}$) in the final states needs to be reversed by appropriately incorporating the charge conjugate matrix $C=-i\gamma^2 \gamma^0$ into the amplitudes. 
This reversed fermion line for the production of diquarks differs from the one for the production of quarkonia by only an additional factor $(-1)^{\zeta+1}$, where $\zeta$ denotes the count of vector vertices along the reversed fermion line \cite{Jiang:2012jt}. 
Then we adopt the following projectors to obtain the (anti-)diquarks for the scalar $[^1S_0]$ state and vector $[^3S_1]$ state,
\begin{equation}
    \begin{split}
&\Pi^{[^1S_0]}_{q_k} = \frac{-\sqrt{M_{k}}}{4 m_{Q_k}m_{Q^\prime_k}}(\slashed{q}_{k2}- m_{Q^\prime_k}) \gamma^5 (\slashed{q}_{k1} + m_{Q_k}),   \\
&\Pi^{[^3S_1]}_{q_k} = \epsilon_\alpha(q_k) \frac{-\sqrt{M_{k}}}{4 m_{Q_k}m_{Q^\prime_k}}(\slashed{q}_{k2}- m_{Q^\prime_k}) \gamma^\alpha (\slashed{q}_{k1} + m_{Q_k}).
\end{split}
\end{equation}
Second, since the colored diquarks can be color-sixtet $\mathbf{6}$ and color-antitriplet $\bar{\mathbf{3}}$, the compact tetraquark has two color configurations 
\begin{equation}
    \begin{split}
& C^{ab;cd}_{\mathbf{6}\otimes\bar{\mathbf{6}}}=\frac{1}{\sqrt{24}}(\delta_{ac}\delta_{bd}+\delta_{ad}\delta_{bc}), \\
& C^{ab;cd}_{\bar{\mathbf{3}}\otimes\mathbf{3}}=\frac{1}{\sqrt{12}}(\delta_{ac}\delta_{bd}-\delta_{ad}\delta_{bc}),
    \end{split}
    \label{eq:compact color}
\end{equation}
where $ab$ and $cd$ are the color indices of the constituent diquarks and the antidiquarks, respectively. 
The scripts in bold are for the color.
$C^{ab;cd}_{\mathbf{6}\otimes\bar{\mathbf{6}}}$ indicates that the constituent diquark is the color-sixtet $\mathbf{6}$ representation and antidiquark is the color-antisixtet $\bar{\mathbf{6}}$ representation, 
and $C^{ab;cd}_{\bar{\mathbf{3}}\otimes\mathbf{3}}$ indicates the constituent diquark is the color-antitriplet $\bar{\mathbf{3}}$ representation and antidiquark is the color-triplet $\mathbf{3}$ representation.
Note that internal diquarks of compact tetraquark can be in either $(QQ)[^3S_1]^{\bar{\mathbf{3}}}$ or $(QQ)[^1S_0]^{\mathbf{6}}$ for $(QQ)$ diquark state with $Q=b,c$ quarks, while for $(bc)$ diquark, they can be in $(bc)[^3S_1]^{\bar{\mathbf{3}}}$, $(bc)[^3S_1]^{\mathbf{6}}$, $(bc)[^1S_0]^{\bar{\mathbf{3}}}$ or $(bc)[^1S_0]^{\mathbf{6}}$ states.

In the two-step transition scheme for the four heavy quarks into the compact tetraquark, the long-distance matrix element ${\langle O_{\overline{{\mathbf{c}}} , \mathbf{c}}^J (T_{C}) \rangle}$ in Eq. \eqref{cross section} can be replaced by two Schr{$\ddot{\mathrm{o}}$}dinger wave functions $\Psi(0)$ at origin for $S$-wave constituent (anti-)diquarks, and $\langle O_{T_C} \rangle$ for diquark and antidiquark being bound into the compact state,
\begin{equation}
\begin{split}
{\langle O_{\overline{{\mathbf{c}}} , \mathbf{c}}^J (T_{C}) \rangle} &\longrightarrow |\Psi_{d_1}(0)|^2 |\Psi_{d_2}(0)|^2 \times \langle O_{T_C} \rangle \\
&=|\Psi_{d_1}(0)|^2 |\Psi_{d_2}(0)|^2 \times \frac{2}{M}|\Psi_{C}(0)|^2 \Gamma^{J} C_{\overline{\mathbf{c}} \otimes \mathbf{c}}^{a b ; c d}\\
&= \frac{1}{(4\pi)^2} \big|R_{d_1}(0)\big|^2 \big|R_{d_2}(0)\big|^2 \times \frac{2}{4\pi M} \big|R_{C}(0)\big|^2 \Gamma^{J} C_{\overline{\mathbf{c}} \otimes \mathbf{c}}^{a b ; c d},
\end{split}
\label{eq:compact transition LDME}
\end{equation}
where the subscripts $d_{1,2}$ and $C$ indicate the two constituent (anti-)diquarks and compact tetraquark respectively, $\Gamma^J$ stands for the polarization summation in Eq. \eqref{summation of polariazations 0,1,2}, and $C_{\overline{\mathbf{c}} \otimes \mathbf{c}}^{a b ; c d}$ is the color factor in Eq. \eqref{eq:compact color}.
$\Psi(0)$ and $R(0)$ are the Schr{$\ddot{\mathrm{o}}$}dinger wave function at origin and the radial wave function at the origin in potential models for $S$-wave states. It is worth noting that, in NRQCD there is no good definition for the long-distance matrix elements for (anti-)diquarks because the (anti-)diquarks are not color-singlet.

With the equations given above, the differential cross section for the production of compact tetraquark is
\begin{equation}
    \begin{split}
        &\mathrm{d}\sigma(e^{+}+e^{-}\to T_C(J)+\gamma) =  \\
        &\mathrm{d}\hat\sigma(e^{+}+e^{-} \to (Q_1Q_2)[n]+(\bar{Q}_2\bar{Q}_1)[n]+\gamma) \times \frac{2 \big|R_{d_1}(0)\big|^2 \big|R_{d_2}(0)\big|^2 \big|R_{C}(0)\big|^2}{(4\pi)^3 M} \ . 
    \end{split}
\end{equation} 
where the polarization summation $\Gamma^J$ and color factor $C_{\overline{\mathbf{c}} \otimes \mathbf{c}}^{a b ; c d}$ have been absorbed into the short-distance cross section $\mathrm{d}\hat\sigma$ which has the dimension of $[M]^{-10}$. 
Note that for the production of $T_C(bb-\bar{b}\bar{b})$ and $T_C(cc-\bar{c}\bar{c})$, we need to multiply an overall factor $\frac{1}{2!2!}$, which accounts for two identical heavy (anti-)quarks inside the component (anti-)diquark.
Here, for antidiquark, we use the same transition probability as its diquark.
In addition, the total angular momentum quantum number of the tetraquark state can take $J=0,1,2$ when $[n]=[^3S_1]$, and $J$ can only be $0$ when $[n]=[^1S_0]$.

\section{NUMERICAL RESULTS}
\label{sec:data}

In this section, we present the numerical results of the cross sections for $e^{+}+e^{-}\to \gamma^{*}/Z^{0}\to T+\gamma$ processes. 
Before the numerical evaluation, some comments should be noted. 
\begin{itemize}
    \item[(a)] Due to the exchange invariance of identical particles, the system composed of two identical particles, i.e. $(c\bar{c})-(c\bar{c})$ and $(b\bar{b})-(b\bar{b})$ here, must satisfy $L+S$=even. So the $S$-wave ($L=0$) molecule-like states $T_M(c\bar{c}-c\bar{c})\,(J=S=1)$ and $T_M(b\bar{b}-b\bar{b})\,(J=S=1)$ do not exist. 
    \item[(b)] Since the constituent quarkonia or (anti-)diquarks are bosons, the $S$-wave molecule-like or compact tetraquarks have the quantum number $P=(-1)^0$ and $C=(-1)^{0+S}$, i.e. $J^{PC} = 0^{++},\, 1^{+-},\, 2^{++}$. So the photon propagated process, $e^{+}+e^{-}\to \gamma^* \to T_{M/C}(J=1)+\gamma$ for $1^{+-}$ molecule-like and compact states are unable to happen because the $C$-parity of two photons is even. 
    \item[(c)] The color factor $C^{ab;cd}_{\mathbf{6}\otimes\bar{\mathbf{6}}}$ is symmetric by the exchange of the two color indices of quarks $a \leftrightarrow b$, however the exchange of the two quarks in Dirac chains will contribute a minus sign because quarks are fermions. So the sum of such two amplitudes for the production of the compact states $T_C((cc)[^1S_0]^{\mathbf{6}}-(\bar{c}\bar{c})[^1S_0]^{\bar{\mathbf{6}}}))$ or $T_C((bb)[^1S_0]^{\mathbf{6}}-(\bar{b}\bar{b})[^1S_0]^{\bar{\mathbf{6}}}))$ is zero.
\end{itemize}

\subsection{Input paramaters} 

We set $m_c=1.5\ {\rm GeV}$ and $m_b=4.9\ {\rm GeV}$. The strong coupling constant is set to be $\alpha_s(2m_c)=0.2355$ for the production of $T(cc\bar{c}\bar{c})$ and $T(bc\bar{b}\bar{c})$, and $\alpha_s(2m_b)=0.1768$ for the $T(bb\bar{b}\bar{b})$ case \cite{Chetyrkin:2000yt}. 
The radial wave functions at the origin for diquarks $|R_d(0)|^2$ are taken from Ref. \cite{Kiselev:2002iy}, and those for quarkona $|R_m(0)|^2$ are taken from Ref. \cite{Yang:2022yxb},
\begin{align}
&|R_{cc}(0)|^2=0.274\ {\rm GeV}^{3}, \quad |R_{bb}(0)|^2=1.810\ {\rm GeV}^{3}, \quad |R_{bc}(0)|^2=0.521\ {\rm GeV}^{3};   \nonumber \\
&|R_{c\bar{c}}(0)|^2=0.528\ {\rm GeV}^{3}, \quad |R_{b\bar{b}}(0)|^2=5.22\ {\rm GeV}^{3}, \  \quad |R_{b\bar{c}}(0)|^2=1.642\  {\rm GeV}^{3}.
\end{align}

For the radial wave functions $R_{M/C}(0)$ at origin for tetraquarks, they can be obtained by numerically solving the Schr$\ddot{\mathrm{o}}$dinger equation under the nonrelativistic potential models. 
The method and the codes are described in Ref. \cite{Lucha:1998xc}.
To evaluate the value of $R_{C}(0)$ for compact tetraquarks, we use a Cornell-inspired potential \cite{Debastiani:2017msn}. \footnote{The Cornell-inspired potential given in Ref. \cite{Debastiani:2017msn} is $V^{(0)}(r)=\kappa_s \frac{\alpha_s}{r}+br$, where $\kappa_s$ is called the ``color factor'' which is related to the color configuration of the system, and $b$ is called ``string tension'' which is related to the strength of the confinement. For an estimate, we take $\kappa_s=-\frac{4}{3}$, $b=0.1463$ for $[n]=[^3S_1]$ and $\kappa_s=-\frac{10}{3}$, $b=1.463$ for $[n]=[^1S_0]$.} And the numerical results are
\begin{align}
&|R_{C}(0)|^2((cc)[^3S_1]^{\bar{\mathbf{3}}}-(\bar{c}\bar{c})[^3S_1]^{\mathbf{3}})=8.422\ {\rm GeV}^3, \nonumber \\
&|R_{C}(0)|^2((bb)[^3S_1]^{\bar{\mathbf{3}}}-(\bar{b}\bar{b})[^3S_1]^{\mathbf{3}})=11.739\ {\rm GeV}^3,\nonumber  \\
&|R_{C}(0)|^2((bc)[^3S_1]^{\bar{\mathbf{3}}/\mathbf{6}}-(\bar{b}\bar{c})[^3S_1]^{\mathbf{3}/\bar{\mathbf{6}}})=7.708\ {\rm GeV}^3, \nonumber \\
&|R_{C}(0)|^2((bc)[^1S_0]^{\bar{\mathbf{3}}/\mathbf{6}}-(\bar{b}\bar{c})[^1S_0]^{\mathbf{3}/\bar{\mathbf{6}}})=74.880\ {\rm GeV}^3.  
\label{Cornell potential1}
\end{align}
Here we adopt the same radial wave functions for the color (anti-)triplet diquark pairs and the color (anti-)sextet diquark pairs for the compact tetraquark $T_C(bc-\bar{b}\bar{c})$ as a rough estimate.
Regarding $R_{M}(0)$ for molecule-like tetraquarks, there is no targeted potential models at present, and we also use the Cornell-like potential to make a rough estimation \footnote{We use the Cornell-inspired potential given in Ref. \cite{Debastiani:2017msn}, i.e. $V^{(0)}(r)=\kappa_s \frac{\alpha_s}{r}+br$, and we take $\kappa_s=0$ and $b=0.1463$ for an estimate because the constituent quarkonium in tetraquark is color singlet.}, and the results are
\begin{align}
&|R_{M}(0)|^2((c\bar{c})[^3S_1]-(c\bar{c})[^3S_1])=0.439\ {\rm GeV}^3,   \nonumber \\
&|R_{M}(0)|^2((b\bar{b})[^3S_1]-(b\bar{b})[^3S_1])=1.434\ {\rm GeV}^3,   \nonumber  \\
&|R_{M}(0)|^2((b\bar{c})[^3S_1]-(c\bar{b})[^3S_1])=0.936\ {\rm GeV}^3.
\label{Cornell potential2}
\end{align}

\subsection{Production of tetraquark at B factory}

In this subsection, we focus on the production of fully heavy tetraquarks at B factories, where the dominant contribution arises from the $\gamma$ mediator, while the contribution from the $Z$-boson mediator is neglected. 
At B factories, we set $\sqrt{s} = 10.58\ {\rm GeV}$ and $\alpha_{\text{EM}}(10.58\ \text{GeV}) = 1/130.9$. Due to the constraint of the center-of-mass energy, only $T(cc\bar{c}\bar{c})$ can be produced at B factories. The production cross sections of the molecular-like tetraquark $T_M(c\bar{c}-c\bar{c})$ and compact tetraquark $T_C(cc-\bar{c}\bar{c})$ in association with a photon are presented in Table \ref{B factory}.
For $T_M(c\bar{c}-c\bar{c})$, we only consider constituent quarkonia in the $^3S_1$ state (i.e., $J/\psi$), and the $J^{PC}$ assignments of the $S$-wave $T_M(c\bar{c}-c\bar{c})$ are $0^{++}$ and $2^{++}$. For $T_C(cc-\bar{c}\bar{c})$, the constituent (anti-)diquarks can only be in the $^3S_1$ state (as per Comment (c) discussed earlier), and the $J^{PC}$ assignments of the $S$-wave $T_C(cc-\bar{c}\bar{c})$ are $0^{++}$ and $2^{++}$.

\begin{table}[ht]
    \caption{Cross sections for the production of $T_M(c\bar{c}-c\bar{c})$ and $T_C(cc-\bar{c}\bar{c})$ in association with a photon at the B factory. Note, the component quarkonia and (anti-)diquarks are both in $[^3S_1]$ states here.}
    \begin{center}
       \begin{tabular}{c|c|c}
        \toprule
        \hline
            $\sigma\ (\mathrm{in}\ ab)$  & $J=0$  &  $J=2$  \\
        \hline
        $T_M((c\bar{c}))[^3S_1]-(c\bar{c}))[^3S_1])$ &
        $0.74$ &
        $0.25$ \\
        $T_C((cc)[^3S_1]^{\bar{\mathbf{3}}}-(\bar{c}\bar{c})[^3S_1]^{\mathbf{3}})$ &
        $2.62$ &
        $20.45$ \\
        \botrule
      \end{tabular}
    \end{center}
    \label{B factory}
\end{table}

Comparing our results with those in Ref. \cite{Feng:2020qee}, where $T_C((cc)[^3S_1]^{\bar{\mathbf{3}}}-(\bar{c}\bar{c})[^3S_1]^{\mathbf{3}})$ with $J=0,2$ are estimated, we find consistent results using the same input parameters.
The target integrated luminosity of the Belle II experiment is $50\ \text{ab}^{-1}$. Further considering the 12\% branching fraction for the leptonic decay of $J/\psi$ and the detection efficiency, observing any fully charmed tetraquark events at Belle II remains pessimistic.
We find that the cross sections for compact states are significantly larger than those for molecular-like states with the same spin, and the cross section for the $2^{++}$ $T_C(cc-\bar{c}\bar{c})$ state is roughly one order of magnitude greater than that of the $0^{++}$ state. 

\begin{figure}
    \centering
    \includegraphics[width=0.49\linewidth]{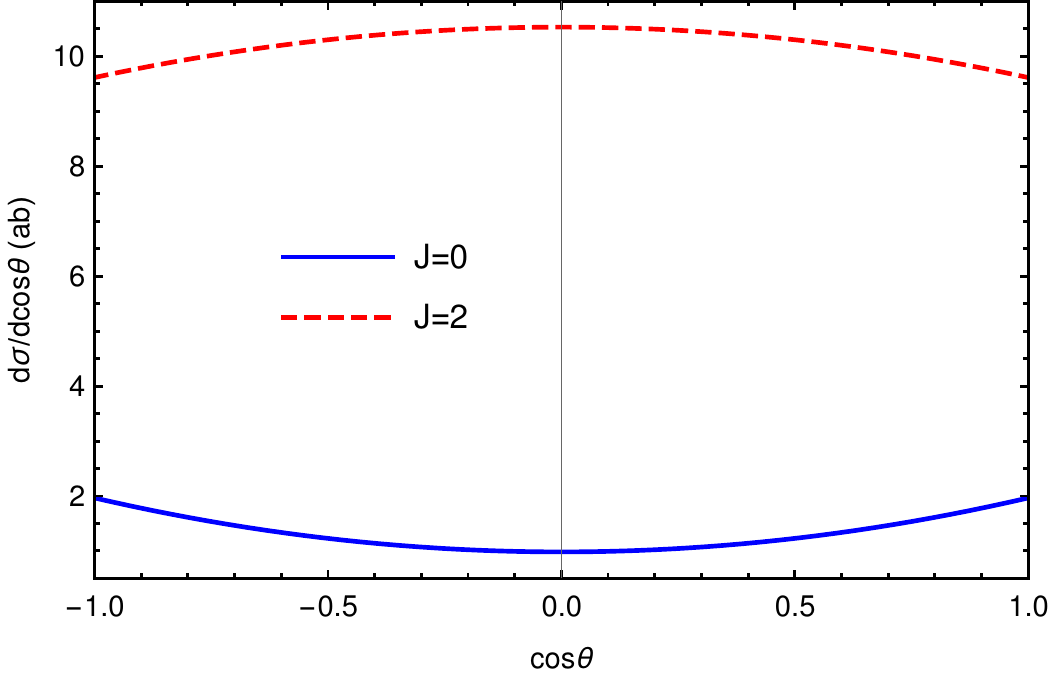}
    \includegraphics[width=0.49\linewidth]{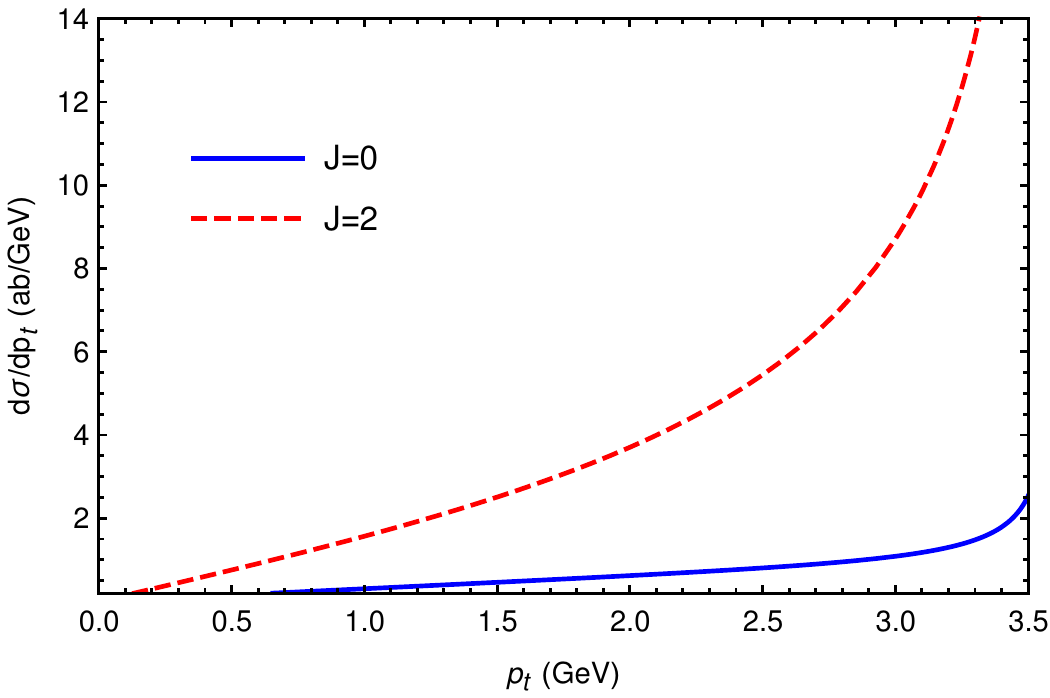}
    \caption{The differential distributions $\mathrm{d}\sigma/\mathrm{d} cos\theta$ and $\mathrm{d}\sigma/\mathrm{d} p_t$ for the process $e^{+}+e^{-} \to T_C((cc)[^3S_1]^{\bar{\mathbf{3}}}-(\bar{c}\bar{c})[^3S_1]^{\mathbf{3}})(J)+\gamma$ with $J=0, 2$ at the B factory, where $\theta$ is the angle between the tetraquark and the beam and $p_t$ is the transverse momentum of the tetraquark.}
    \label{fig:cospt}
\end{figure}

In Fig. \ref{fig:cospt}, we present the differential distributions for the process $e^{+}+e^{-} \to T_C((cc)[^3S_1]^{\bar{\mathbf{3}}}-(\bar{c}\bar{c})[^3S_1]^{\mathbf{3}})(J=0,2)+\gamma$ versus $\cos\theta$ and $p_t$ at B factories, where $\theta$ denotes the angle between the tetraquark and the beam, and $p_t$ is the transverse momentum of the tetraquark.
The lower solid blue lines correspond to the $0^{++}$ $T_C((cc)[^3S_1]^{\bar{\mathbf{3}}}-(\bar{c}\bar{c})[^3S_1]^{\mathbf{3}})$ state, while the upper dashed red lines correspond to the $2^{++}$ state. For the angular distribution, the $2^{++}$ $T_C((cc)[^3S_1]^{\bar{\mathbf{3}}}-(\bar{c}\bar{c})[^3S_1]^{\mathbf{3}})$ state exhibits a symmetric "bulge" profile, whereas the $0^{++}$ state shows a symmetric "hollow" profile. The angular distribution of the unpolarized $J=2$ compact state is consistent with that reported in Ref. \cite{Feng:2020qee}.
In comparison with the $0^{++}$ state, the transverse momentum distribution of the $2^{++}$ state increases considerably more rapidly.

\subsection{Production of tetraquark at Z factory}

In this subsection, we focus on the production of fully heavy tetraquarks when the center-of-mass energy is at the $Z$ boson pole, where the contribution from the $\gamma$ mediator is neglected. We set $\sqrt{s} = 91.2\ {\rm GeV}$, and the other parameters adopt the following values: $\sin^2\theta_W = 0.23119$, the $Z$ boson mass $m_Z = 91.1876\ {\rm GeV}$, and its total decay width $\Gamma_Z = 2.4952\ {\rm GeV}$ \cite{ParticleDataGroup:2018ovx}. The production cross sections for molecular-like and compact tetraquarks are presented in Table \ref{Z factory molecule} and Table \ref{Z factory compact}, respectively.
For molecular-like tetraquarks ($T_M$), we only consider constituent quarkonia in the $^3S_1$ state. For $T_{M}(c\bar{c}-c\bar{c})$ and $T_{M}(b\bar{b}-b\bar{b})$, the possible $J^{PC}$ assignments are $0^{++}$ and $2^{++}$, as the $J=1$ configuration is excluded by the constraint in Comment (a). For $T_{M}(b\bar{c}-\bar{b}c)$, the allowed $J^{PC}$ assignments are $0^{++}$, $1^{+-}$, and $2^{++}$.
For compact tetraquarks ($T_C$), the constituent (anti-)diquarks can only be in the $^3S_1$ state for $T_{C}(cc-\bar{c}\bar{c})$ and $T_{C}(bb-\bar{b}\bar{b})$ (due to Comment (c), which does not apply to $T_{C}(bc-\bar{b}\bar{c})$). For $T_C$, the $J^{PC}$ assignments are $0^{++}$, $1^{+-}$, and $2^{++}$, with no constraints from Comments (a, b).

\begin{table}[ht]
    \caption{Cross sections for the production of molecule-like tetraquark $T_M(c\bar{c}-c\bar{c})$, $T_M(b\bar{b}-b\bar{b})$ and $T_M(b\bar{c}-\bar{b}c)$ at the Z factory. Note, the constituent quarkonia are all in $[^3S_1]$ states.}
   \begin{center}
       \begin{tabular}{c|c|c|c}
        \toprule
        \hline
            $\sigma\ (\mathrm{in}\ ab)$  & $J=0$  &  $J=1$ &  $J=2$  \\
        \hline
        $T_M((c\bar{c})[^3S_1]-(c\bar{c})[^3S_1])$ & $0.0093$ & / & $0.012$ \\
        $T_M((b\bar{b})[^3S_1]-(b\bar{b})[^3S_1])$ & $0.00034$ & / &$0.00050$ \\
        $T_M((b\bar{c})[^3S_1]-(\bar{b}c)[^3S_1])$ & $0.0045$ & $0.0020$ & $0.011$ \\
        \botrule
      \end{tabular}
    \end{center}
    \label{Z factory molecule}
\end{table}

\begin{table}[ht]
    \caption{Cross sections for the production of compact tetraquark $T_C(bb-\bar{b}\bar{b})$, $T_C(cc-\bar{c}\bar{c})$ and $T_C(bc-\bar{b}\bar{c})$ at the Z factory. Here $\sigma [T_C(bc-\bar{b}\bar{c})]=\sigma [T_C((bc)[^1S_0]^{\mathbf{6}}-(\bar{b}\bar{c})[^1S_0]^{\bar{\mathbf{6}}})] + \sigma [T_C((bc)[^1S_0]^{\bar{\mathbf{3}}}-(\bar{b}\bar{c})[^1S_0]^{\mathbf{3}})]+\sigma [T_C((bc)[^3S_1]^\mathbf{6} -(\bar{b}\bar{c})[^3S_1]^{\bar{\mathbf{6}}})]+\sigma [T_C((bc)[^3S_1]^{\bar{\mathbf{3}}}-(\bar{b}\bar{c})[^3S_1]^{\mathbf{3}})]$.}
   \begin{center}
       \begin{tabular}{c|c|c|c}
        \toprule
        \hline
            $\sigma\ (\mathrm{in}\ ab)$  & $J=0$  &  $J=1$ &  $J=2$  \\
        \hline
        $T_C((cc)[^3S_1]^{\bar{\mathbf{3}}}-(\bar{c}\bar{c})[^3S_1]^{\mathbf{3}})$ &
        $0.071$ &
        $1.06$ &
        $0.093$ \\
        $T_C((bb)[^3S_1]^{\bar{\mathbf{3}}}-(\bar{b}\bar{b})[^3S_1]^{\mathbf{3}})$ &
        $0.024$ &
        $0.0022$ &
        $0.092$ \\
        $T_C(bc-\bar{b}\bar{c})$ &
        $0.13$ &
        $0.039$ &
        $0.0099$ \\
        \botrule
      \end{tabular}
    \end{center}
    \label{Z factory compact}
\end{table}

Although the designed integrated luminosity of CEPC in GigaZ mode is as high as approximately $16\ \text{ab}^{-1}$, and that of FCC-ee is roughly nine times that of CEPC, the detection of both molecular-like and compact tetraquarks remains nearly impossible due to the extremely small number of events.
For $T_C((cc)[^3S_1]^{\bar{\mathbf{3}}}-(\bar{c}\bar{c})[^3S_1]^{\mathbf{3}})$, it is noteworthy that the contribution of the $J=1$ state is more than one order of magnitude larger than those of the $J=0$ or $J=2$ states.
For $T_C((bb)[^3S_1]^{\bar{\mathbf{3}}}-(\bar{b}\bar{b})[^3S_1]^{\mathbf{3}})$, the contribution of the $J=2$ state is approximately 42 times that of the $J=1$ state.
For $T_C(bc-\bar{b}\bar{c})$, the contribution of the $J=0$ state is roughly 13 times that of the $J=2$ state.

\section{SUMMARY} 
\label{sec:summary}

In this work, within the NRQCD framework, we investigate the processes $e^{+}+e^{-}\to \gamma^{*}/Z^{0}\to T+\gamma$, where $T$ denotes the fully heavy tetraquarks $T(bb\bar{b}\bar{b})$, $T(cc\bar{c}\bar{c})$, and $T(bc\bar{b}\bar{c})$. Both molecular-like and compact states with total spin $J=0,1,2$ are considered. We discuss the production rates of these fully heavy tetraquarks at B factories and future $Z$ factories. We find that observing events of any fully heavy tetraquark via the exclusive $e^{+}+e^{-}\to T+\gamma$ process—whether for compact or molecular-like states—is pessimistic, both at Belle II and at future $Z$ factories such as CEPC and FCC-ee.
Nevertheless, we emphasize that our estimation of the NRQCD long-distance matrix element values, which is based on potential models, is approximate. To yield an accurate prediction, more reliable evaluations of all the NRQCD matrix elements encountered in this work are necessary.

\vspace{0.5cm} {\bf Acknowledgments}
This work is supported in part by National Natural Science Foundation of China under the grants 
No. 12235008, No. 12321005, No. 12275157, No. 12475083, No. 12475143. 


\end{document}